\documentstyle[12pt]{article}
\begin{document}
\rightline{DTP-95/43 }
\rightline{hep-th/9512011}

\begin{center}
{\Large On the sine-Gordon---Thirring equivalence\break 
in the presence of
a boundary} 
\vskip 1.0cm
\centerline{Z-M Sheng$^{a,b}$\footnote{e-mail : Sheng.zheng-mao@durham.ac.uk}, 
 H-B Gao$^{a,c}$ } 
\vskip 0.7cm
{ \it a. Department of Mathematical Science, University of Durham, \\[3mm] 
~~~~~Durham, DH1 3LE, U.K. 
\\[3mm]
 b. Department of Physics, Hangzhou University,\\[3mm]
~~~~~Hangzhou 310028, P.R. China \footnote{Permanent address, mail 
address since 10 Jan. 1996}\\[3mm]
 c. Zhejiang Institute of Modern Physics, Zhejiang University,\\[3mm]
~~~~~Hangzhou 310027, P.R. China }\footnote{Permanent address}\\[3mm]
\vskip 0.8cm
{\sc Abstract }
\vskip 0.5cm

\begin{minipage}{12.5cm}
In this paper, the relationship between the sine-Gordon model with an 
integrable 
boundary condition and  the Thirring model with boundary is discussed and 
the reflection   $R$-matrix for the massive Thirring model, which 
is related to the physical boundary parameters of 
the sine-Gordon model,  is given. The relationship between the 
the boundary parameters and the two formal parameters appearing in the 
work of Ghoshal and Zamolodchikov  is discussed. 
\end{minipage}
\end{center}
\vskip 0.2cm
\noindent November 1995
\vfill
\eject

\noindent{\bf Introduction}

Recently, much attention has been paid to the study of integrable 
field theory with 
boundary  \cite{GZ}\cite{CDRS}\cite{BCDR}\cite{Mac}\cite{Gh}\cite{SS}\cite{PZ}. 
The study of such  field theories is not only  intrinsically interesting 
 but also provides a better understanding of boundary related 
phenomena in statistical physics and condensed matter  \cite{LeC}\cite{Wen}. 
Probably the most famous physical example of a boundary integrable model 
is the Kondo problem, where the 1+1 dimensional field theory is an effective 
field theory of s-wave scattering of electrons off a magnetic spin 
impurity. Such an impurity problem, in which one concentrates on s-wave 
scattering from some isolated object at the origin, generically provides 
interesting 1+1 dimensional boundary field theory. There are also 
1+1 dimensional boundary quantum integrable systems of 
experimental relevance, such as Luttinger liquid (Thirring) model for 
the edge states of electrons in the fractional quantum Hall 
effect  \cite{Wen} and the boundary sine-Gordon model may provide an 
accurate description of conductance through a point contact  \cite{FLS}.

An integrable field theory possesses an infinite set of independent, 
commuting integrals of 
motion. In the \lq bulk theory' these integrals of motion follow from an 
infinite number
of conserved currents. However, when the theory is restricted to a half-line
(or to an interval) the 
existence of conserved currents on the whole-line does not 
guarantee integrability unless special boundary conditions are specified. 
The integrable boundary conditions under which 
a theory preserves its integrability can be obtained in various
ways: by the boundary 
action functional, via a perturbed conformal boundary 
condition  \cite{Cardy}\cite{CL} or through a Lax pair approach  \cite{BCDR}.

An important characteristic of an integrable field theory on the whole-line
is its 
factorizable   $S$-matrix. In the bulk theory, such an $S$-matrix is 
required to 
satisfy the Yang-Baxter equation (or \lq  factorizability 
condition'), in addition to the standard requirements  of unitarity 
and crossing. 
These equations have much restrictive power, determining the $S$-matrix up 
to the so-called  \lq  CDD ambiguity' \cite{AK}\cite{ZZ}\cite{Kor}. 
For an integrable field theory  with a
boundary, particles cannot escape beyond the boundary and therefore
reflect from it. The assumption of factorisibility for a theory defined 
on a half-line requires the factorizable    $S$-matrix describing the 
scattering of particles in the bulk far from the boundary to be 
compatible with the
matrix describing the reflection from the boundary (the 
boundary reflection matrix often denoted $R$ or $K$).
$S$ and $R$ are required to satisfy an 
appropriate generalization of the Yang-Baxter equation (the Boundary 
Yang-Baxter equation \cite{GZ}), and also generalizations of 
unitarity and crossing (the Boundary Cross-Unitarity 
equation \cite{GZ}\cite{FS}). There is also a modified version of the 
bootstrap equations (the Boundary Bootstrap \cite{FK}\cite{Sasaki}) which 
ensures
compatibility between the $R,S$ descriptions of particles and their
bound states. 

A particularly interesting model of massive boundary field theory is 
the sine-Gordon model on a half line. It exhibits relationships with 
the theory 
of Jack Symmetric functions  \cite{FLS}, and has applications to 
dissipative quantum mechanics \cite{GHM} and impurity problems in 
a one-dimensional 
strongly correlated electron gas  \cite{KF}. Including the 
integrable 
boundary condition, its action can be written  \cite{GZ}
\begin{equation}
\begin{array}{ll}
S=& \int^{+\infty}_{-\infty}dt
\int^{0}_{-\infty}dx\left[\frac{1}{2}(\partial_x\phi)^{2}
+\frac{1}{2}(\partial_t\phi)^{2}-\frac{m_0^2}{\beta^2}\cos\beta\phi 
-\frac{m_0^2}{\beta^2}\right] \\[3mm]
& - h\int^{+\infty}_{-\infty}dt\left(\cos\frac{\beta(\phi-\phi_0)}{2}
\Bigm|_{x=0}
-\cos \frac{\beta\phi_0}{2}\right),
\end{array}
\end{equation}
where $\phi(x,t)$ is a real scalar field, $\beta$ is a dimensionless
coupling constant and $m_0$ is the mass parameter, the constant term is 
added so that the energy is zero when $\phi=0$. Its integrable boundary 
condition reads
\begin{equation}
\partial_x\phi = \frac{\beta}{2}h\sin\frac{\beta}{2}
\left(\phi-\phi_0\right) 
~~~~~~\hbox{at} ~~ x=0,  
\end{equation} 
and it tends to a free boundary condition (or  fixed boundary condition), 
when $h \rightarrow 0 $ (or $h \rightarrow \infty$ ).
The $R$ matrix for this model (modulo the  \lq CDD ambiguity') was 
obtained by explicitly solving the 
boundary Yang-Baxter equation and the crossing-unitarity equations   
\cite{GZ}. This general solution 
to the boundary Y-B equation and the C-U condition depends on two formal 
parameters. However the relationship between  these
formal parameters and the physical parameters ($h$ and $\phi_0$) 
related to the boundary term in the action
were not given. There are also other methods
for obtaining the $R$ matrix, such as via 
the Bethe Ansatz  \cite{FS}\cite{SS} or from 
perturbed conformal field theory  \cite{AKL}. However, until now 
only the $R$-matrices for special 
boundary conditions, such as the free boundary condition, the fixed boundary 
condition and the free fermion point, were given. The study of how
the $R$-matrix 
is related to the 
physical boundary parameters for the sine-Gordon model with a general 
integrable boundary 
condition remains to be undertaken. In this paper, we intend to improve 
this situation by
 giving such a $R$-matrix, making use of the well-known 
relationship between the Thirring Model and the sine-Gordon 
model \cite{Col}\cite{Ma}. 

One can construct a transformation 
to fermionize the bulk sine-Gordon model into the bulk massive Thirring 
model  \cite{Ma} in terms of the non-local transformation recently used  in 
a discussion of the non-local currents of the sine-Gordon 
model  \cite{BL}\cite{CR}\cite{YC}\cite{YLS}. However, there are several kinds 
of integrable boundary conditions for the Thirring Model (which
is therefore rather different to the 
sine-Gordon model in which there is only one class of integrable 
boundary condition). For example, there are $SU(2)$ invariant or $U(1)$
 invariant 
integrable boundary 
conditions  \cite{Luk}\cite{CHSWY}\cite{SS}; the  richer 
boundary structure in the Thirring model matches the situation  in 
the bulk case  \cite{KN}\cite{KM}. If we wish, we can choose a 
boundary condition  
equivalent to  a boundary condition of the sine-Gordon model but which
 is linear in terms  of the Thirring field. In the third section, we 
will give the $R$-matrix of the sine-Gordon model obtained
using a special integrable 
boundary condition for the massive Thirring model. 

\noindent{\bf Relation between sine-Gordon model and Thirring model 
with boundary}

In order to fermionize the sine-Gordon model, the following 
non-local transformation may be introduced:
\begin{equation}
\left \{
\begin{array}{ll}
\rho(x,t)&=\frac{1}{2}\left(\phi(x,t) + 
\int^{x}_{-\infty}\partial_t{\phi}(y,t)dy \right) 
\\[3.mm]
\bar{\rho}(x,t) &=\frac{1}{2}\left(\phi(x,t) -  
\int^{x}_{-\infty}\partial_t{\phi}(y,t)dy\right).
\end{array} 
\right.
\end{equation}
Using this transformation, we define new fields by
\begin{equation}
\begin{array}{lll}
\psi_1(x)&=A:e^{-ia\rho(x)-ib\bar{\rho}(x)}: ~~~~~~~~~~
\psi_1^{\dagger}(x)& = A:e^{ia\rho(x)+ib\bar{\rho}(x)}:\\[3mm]
\psi_2(x) &= -iA :e^{ib\rho(x)+ia\bar{\rho}(x)}:~~~~~~~~~
\psi_2^\dagger(x)& = iA:e^{-ib\rho(x)-ia\bar{\rho}(x)}:
\end{array}
\end{equation}
where $a = \frac{1}{2}(\beta+\frac{4\pi}{\beta})$, 
~~$b=\frac{1}{2}(\beta-\frac{4\pi}{\beta})$, and $A$ is a 
constant with the dimension $[M]^{\frac{1}{2}}$.
It is easy to compute the anticommutators for the fields $\psi_i(x)$ 
\begin{equation}
\{ \psi_i(x,t), \psi^{\dagger}_j(y,t) \} = \delta_{ij}\delta(x-y)
\end{equation}
by using the  
canonical  commutation relations for the field $\phi(x,t)$ and its
canonical conjugate $\partial_t\phi(y,t)$
\begin{equation}
[ \phi(x,t), \partial_t\phi(y,t)] = i\delta(x-y),
\end{equation}
and the standard relation:
\begin{equation}
e^{A}e^{B} = e^{[A,B]}e^Be^A,
\end{equation}
valid when $[A,B]$ commutes with both $A$ and $B$.

Using the above transformation, one can obtain the Lagrangian of 
the bulk Thirring 
Model as follows \cite{Col}\cite{Ma}
\begin{equation}
{\cal L}_{\rm T_{\rm bulk}} = \bar{\psi} i \gamma^{\mu}
\partial_{\mu}\psi - \frac{g}{2}j^\mu 
j_{\mu} - M\bar{\psi}\psi,
 \end{equation}
where $\psi= \left (
\begin{array}{c}
\psi_1 \\[3mm]
\psi_2 
\end{array}
\right )$ and $\bar{\psi}=\psi^\dagger \gamma^0 = (\psi_2^\dagger , 
\psi_1^\dagger  ) $ with
$$
\gamma^1 = \left (\begin{array}{cc}
 0 & -1\\
1 & 0 \end{array} \right )
 ~~~~~
\gamma^0 = \left (\begin{array}{cc}
0 & 1\\
1 & 0 \end{array} \right )
$$
and the current components and the various parameters defined by:
\begin{equation}
\left \{
\begin{array}{ll}
& j^{\mu}(x) = \lim\limits_{y\rightarrow 
x}\frac{1}{2}\left(\bar{\psi}(x)\gamma^{\mu}\psi(y) + 
\bar{\psi}(y)\gamma^\mu\psi(x)\right) = 
-\frac{\beta}{2\pi}\epsilon^{\mu\nu}\partial_{\nu}\phi(x)\\[3mm]
 &\frac{4\pi}{\beta^2} = 1 + g/\pi \\[3mm]
 &\frac{m_0^2}{\beta^2}\cos \beta\phi =  Z M \bar{\psi}\psi \end{array}
\right.,
\end{equation}
here $ \epsilon^{01} = 1$ and $ Z= \frac{m_0^2}{2 M\beta^2 A^2} $. From 
the first of eqs(9)\footnote{This form of $j^\mu$ was taken
in \cite{Som} to avoid the singularity}, we know that $j^0(x) = 0 $ at $ 
x=0$ corresponds to the free boundary condition of the sine-Gordon model, 
and $j^1(x) = 0$ at $ 
x=0 $ corresponds to $\partial_t\phi(0,t) = 0 $ or $\phi(0,t) = \phi_0 $
which is  the fixed boundary condition. Explicitly,
\begin{equation}
j^0(x, t)\Bigm|_{x=0} = 
\frac{1}{2}(\psi_1^\dagger (0)\psi_1(0^-)+\psi_2^\dagger (0)\psi_2(0^-) + 
\psi_1^\dagger (0^-)\psi_1(0) + \psi_2^\dagger (0^-)\psi_2(0)) = 0
\end{equation}
or
\begin{equation}
j^1(x, t)\Bigm|_{x=0} = 
\frac{1}{2}(\psi_1^\dagger (0)\psi_1(0^-) - \psi_2^\dagger (0)\psi_2(0^-) + 
\psi_1^\dagger (0^-)\psi_1(0) - \psi_2^\dagger (0^-)\psi_2(0)) = 0,
\end{equation}
which imply the components $\psi_1(x)$ and $ \psi_2(x)$ must be dependent
at $x=0$.
Note also, if
\begin{equation}
\left \{
\begin{array}{ll}
\psi_1^\dagger  (0) & = \mu\psi_2(0) \\[3mm]
\psi_2^\dagger (0) & = \mu\psi_1 (0),
\end{array}
\right.
\end{equation}
where $\mu$ is a phase commuting with the 
components of $\psi$, then $j^0(0) =0 $  
 corresponding to the free boundary condition. On the other hand, if 
\begin{equation}
\left \{
\begin{array}{ll}
\psi_1 (0) & = e^{-i\beta\phi_0}\psi_2(0) \\[3mm]
\psi_1^\dagger (0) & = e^{i\beta\phi_0}\psi_2^\dagger(0),
\end{array}
\right.
\end{equation}
then $j^1(0) = 0$ which corresponds to the fixed boundary condition 
$$\phi=\phi_0 + \frac{2\pi n}{\beta}.$$

The Thirring model restricted to the half-line with a general boundary 
condition may be thought of as a perturbation of the action
containing the free condition:
\begin{equation}
{\cal L}_{\rm T_{\rm b}} = {\cal L}_{\rm T_{\rm free}} + 
{\cal L}_{\rm T_{\rm boundary}},
\end{equation}
where, the action with free boundary condition  reads, 
\begin{equation}
{\cal L}_{\rm T_{\rm free}}= {\cal L}_{\rm T_{\rm bulk}}+ i\mu 
\left( \psi_2(0)\psi_1^{\phantom{\dagger}}(x) - \psi_1(0)\psi_2(x) 
\right) 
\delta(x)+ i\mu^\dagger \left( \psi_2^\dagger (x)\psi_1^\dagger (0) - 
\psi_1^\dagger (x)\psi_2^\dagger (0)\right)\delta (x).
\end{equation}
(From this action, one can easily obtain the free boundary condition, eq(12).)

Next, we want to find the boundary Lagrangian density corresponding 
to the general sine-Gordon boundary term. Noting,
\begin{equation}
:\cos\beta\left(\frac{\phi-\phi_0}{2}\right) :\ = 
\frac{1}{2} :e^{i\frac{\beta\phi}{2}}:e^{-i\frac{\beta\phi_0}{2}}
+\frac{1}{2} :e^{-i\frac{\beta\phi}{2}}:e^{i\frac{\beta\phi_0}{2}},
\end{equation}
we introduce the zero mode operators (or \lq boundary field 
operators'), $b_\pm$: 
\begin{equation}
\begin{array}{ll}
b_{-} & = \frac{1}{4A} 
\exp\left[-\frac{i\pi}{2}(1+\frac{4}{\beta}\int^0_{-\infty}
\dot{\phi}(\xi,t)d\xi)\right] \\[3mm]
b_+  & = \frac{1}{4A} 
\exp\left[\frac{i\pi}{2}(1+\frac{4}{\beta}\int^0_{-\infty}\dot{\phi}(\xi,t)
d\xi)\right],
\end{array}
\end{equation}
and use them to construct the required term.
Since one can check that $b_\pm$ anticommutate with the components of
$\psi$ and 
$\psi^\dagger $,
eq(16) becomes
\begin{equation}
:\cos\beta\left(\frac{\phi-\phi_0}{2}\right) :\ = 
(\psi_1^\dagger b_{-} +\psi_2 b_+ )e^{-i\frac{\beta\phi_0}{2}}
-(\psi_1b_+  +\psi_2^\dagger  b_{-})e^{i\frac{\beta\phi_0}{2}}
\end{equation}
and the boundary Lagrangian density is
\begin{equation}
\begin{array}{ll}
{\cal L}_{\rm T_{\rm boundary}} =& h \left((\psi_1^\dagger b_{-} - b_+\psi_2 
 ) e^{-i\frac{\beta\phi_0}{2}}
+ (b_+\psi_1 -\psi_2^\dagger  b_{-})e^{i\frac{\beta\phi_0}{2}} - 
\cos\beta\frac{\phi_0}{2}\right)\delta(x) \\[3mm]
&+ i h b_+ \partial_t b_{-}\delta(x). 
\end{array}  
\end{equation}

Varying the action (14) 
with respect to the fermion fields and boundary field 
operator, we get the boundary conditions:
\begin{equation}
\left \{
\begin{array}{ll}
(\psi_1^\dagger  - \mu\psi_2) e^{-i\beta\phi_0} - 
(\psi_2^\dagger  -\mu\psi_1) = 0 \\[3mm]
\partial_t(\psi_1 - \mu^\dagger \psi_2^\dagger ) +  
h(\psi_2e^{-i\beta\phi_0} -\psi_1) = 0  \\[3mm]
\partial_t(\psi_1^\dagger  - \mu\psi_2) -  h(\psi_1^\dagger  - 
\psi_2^\dagger e^{i\beta\phi_0}) = 0.
\end{array}
\right.
\end{equation}
 It is worth noting that while the boundary 
condition (2) is the unique form of integrable boundary condition for 
the sine-Gordon 
model, at least on the assumption its boundary potential 
${\cal B}(\phi)$ only depends on $\phi$,  the boundary 
condition (20) is not the only integrable boundary condition for the massive 
Thirring model even if its boundary potential depends only on  
$\psi_i$ (and not its $x$-derivatives) at $x=0$.

\noindent{\bf $R$ matrix related to the boundary parameters of sine-Gordon}

The bulk theory (8) contains two types of fermion, a \lq soliton' and 
an \lq antisoliton', each of mass $M$. The corresponding particle creation 
operators $A_{-}^\dagger (\theta)$ and $A_+^\dagger (\theta)$ can be defined 
through the decomposition:
\begin{equation}
\left \{
\begin{array}{c}
\psi_1(x) = -i \sqrt{\frac{M}{4\pi}}\int d\theta e^{\theta/2}( 
A_+ (\theta) e^{iMx\sinh\theta - iMt\cosh\theta} -
A_{-}^\dagger (\theta) e^{-iMx\sinh\theta + iMt\cosh\theta}) \\[3mm]
\psi_1^\dagger (x) = -i \sqrt{\frac{M}{4\pi}}\int d\theta e^{\theta/2}( 
A_{-}(\theta) e^{iMx\sinh\theta - iMt\cosh\theta} -
A_+ ^\dagger (\theta) e^{-iMx\sinh\theta + iMt\cosh\theta}) \\[3mm]
\psi_2(x) = - \sqrt{\frac{M}{4\pi}}\int d\theta e^{-\theta/2}( 
A_+ (\theta) e^{iMx\sinh\theta - iMt\cosh\theta} +
A_{-}^\dagger (\theta) e^{-iMx\sinh\theta + iMt\cosh\theta}) \\[3mm]
\psi_2^\dagger (x) = - \sqrt{\frac{M}{4\pi}}\int d\theta e^{-\theta/2}( 
A_{-}(\theta) e^{iMx\sinh\theta - iMt\cosh\theta} +
A_+ ^\dagger (\theta) e^{-iMx\sinh\theta + iMt\cosh\theta}) ,
\end{array}
\right.
\end{equation}
where $\theta$ is the usual rapidity variable, and momentum and energy 
are given by
$$
P = M \sinh\theta, ~~~~~~~E= M\cosh \theta,
$$ 
$M$ is the mass of soliton or antisoliton.
The exact relation between $M$ and $m_0$ is  \cite{Zam}
\begin{equation}
M={\cal 
K}(\beta)m_0^{{8\pi}/{\beta^2\lambda}}\Lambda^{-{1}/{\lambda}},
\end{equation}
where $\Lambda$ is an ultraviolet cutoff, $\lambda=\frac{8\pi}{\beta^2}-1$, and
$$
{\cal K}(\beta) = 
\frac{2\Gamma(\frac{1}{2\lambda})}{\sqrt{\pi}\Gamma(\frac{\lambda+1}
{2\lambda})}\left (  \frac{(\lambda+1)\Gamma(\frac{\lambda}{\lambda+1})}
{16\Gamma(\frac{1}{1+\lambda})} \right ) ^{\frac{\lambda+1}{2\lambda}} . $$
As $\beta \rightarrow 0$, $\lambda \rightarrow \infty $, ${\cal K}(\beta) 
\rightarrow 
\frac{8}{\beta^2} $, one obtains the well known classical expression:
$$
M = \frac{8m_0}{\beta^2}. 
$$ 
 
Using the anti-commutation relations of $\psi_i$, one calculates  the 
anti-commutators for $A$'s to be:
\begin{equation}
\begin{array}{ll}
\{ A_{\pm}(\theta), A_{\pm}^\dagger (\theta^{\prime}) \} &= 
\delta(\theta-\theta^{\prime}) \\[3mm]
\{ A_{\pm}(\theta), A_{\mp}^\dagger (\theta^{\prime}) \} &= 0.
\end{array}
\end{equation}
Substituting eq(21) into the boundary conditions  (20), we find the 
following relations
\begin{equation}
\begin{array}{ll}
&\left(i M\cosh\theta e^{\theta/2} + i h e^{-\theta/2}e^{-i\beta\phi_0} - h 
e^{\theta/2}\right)A_{-}^\dagger (\theta)\, B +\mu^\dagger  M\cosh\theta 
e^{-\theta/2}A_+ ^\dagger (\theta)\, B \\[3mm]
&+ \left(i M\cosh\theta e^{-\theta/2} + ih e^{\theta/2}e^{-i\beta\phi_0} 
- h e^{-\theta/2}\right)A_{-}^\dagger (-\theta)\, B \\[3mm]
&+\mu^\dagger  M\cosh\theta 
e^{\theta/2}A_+ ^\dagger (-\theta)\, B = 0 
\end{array}
\end{equation}
and
\begin{equation}
\begin{array}{ll}
\left(i M\cosh\theta e^{\theta/2} + ih e^{-\theta/2}e^{i\beta\phi_0} - h 
e^{\theta/2}\right)A_+ ^\dagger (\theta)\, B +\mu M\cosh\theta 
e^{-\theta/2}A_{-}^\dagger (\theta)\, B \\[3mm]
+ \left(i M\cosh\theta e^{-\theta/2} + ih e^{\theta/2}e^{i\beta\phi_0} - h 
e^{-\theta/2}\right)A_+ ^\dagger (-\theta)\, B  \\[3mm]
+\mu M\cosh\theta 
e^{\theta/2}A_{-}^\dagger (-\theta)\, B = 0,
\end{array}
\end{equation}
where $B$ represents any boundary state.

Since $B$ represents an arbitrary boundary state, eqs(24,25) imply expressions
for $A_\pm^\dagger (\theta)$ in terms of  $A_\pm^\dagger (-\theta)$ which
may be conveniently written in the form
\begin{equation}
\left ( 
\begin{array}{c}
A_+ ^\dagger (\theta) \\[3mm]
A_{-}^\dagger (\theta) 
\end{array}
\right )B = R(\theta) \left (
\begin{array}{c}
A_+ ^\dagger (-\theta) \\[3mm]
A_{-}^\dagger (-\theta) 
\end{array}
\right ) B
\end{equation}
with the $R$-matrix parametrised by
$$
R(\theta) = \left (
\begin{array}{cc}
P^+ (\theta) & Q^+ (\theta) \\[3mm]
Q^{-}(\theta) & P^{-}(\theta) 
\end{array}
\right ),
$$
where
\begin{equation}
\left \{
\begin{array}{ll}
P^+ (\theta) & = \left(M \cosh\theta + {h}\cosh(\theta + i\beta 
\phi_0)\right) /P(\theta) \\[3mm]
P^{-}(\theta) & =\left(M \cosh\theta + {h}\cosh(\theta - i\beta 
\phi_0)\right) /P(\theta) \\[3mm]
Q^+ (\theta) & = -i{\mu}M\cosh \theta\sinh \theta / P(\theta) \\[3mm]
Q^{-}(\theta) & = - i{\mu^\dagger }M\cosh \theta \sinh2\theta / P(\theta) 
\\[3mm]
P(\theta) & = - M\cosh^2\theta - h(\cos\beta\phi_0 + i\sinh\theta). \end{array}
\right.
\end{equation}
Returning to eq(21), we find  the anticommutators eq(5) and eq(23) do not 
change under the following transformations
\begin{equation}
\left \{
\begin{array}{ll}
A_+ (\theta)& \rightarrow f(\theta) A_+ (\theta) \\[3mm]
A_{-}(\theta)& \rightarrow g(\theta) A_{-}(\theta) 
\end{array}
\right.
\end{equation}
\begin{equation}
\left \{
\begin{array}{ll}
A_+ ^\dagger (\theta)& \rightarrow f^\dagger (\theta) A_+ ^\dagger (\theta) 
\\[3mm]
A_{-}^\dagger (\theta)& \rightarrow g^\dagger (\theta) A_{-}^\dagger 
(\theta), \end{array}
\right.
\end{equation}
where 
$f(\theta)$, $g(\theta)$ are some arbitrary functions of $\theta$ which 
satisfy the following relations
\begin{equation}
f(\theta)f^\dagger (\theta) =1 , ~~~~~~~~g(\theta)g^\dagger (\theta) =1
\end{equation}
In other words, the   $R$-matrix has an additional degree of freedom  
which is similar to the  \lq CDD ambiguity '.
Under the above transformations, the   $R$-matrix becomes
\begin{equation}
R(\theta) \rightarrow R^{'} = \left (
\begin{array}{cc}
f(\theta) & 0 \\[3mm]
0 & g(\theta)
\end{array}
\right ) R(\theta) \left (
\begin{array}{cc}
f^\dagger (-\theta) & 0 \\[3mm]
0 & g^\dagger (-\theta)
\end{array}
\right )
\end{equation}
or
$$
\left \{
\begin{array}{ll}
& P^{'+}(\theta) = f(\theta)f^\dagger (-\theta) P^\dagger (\theta) \\[3mm]
& P^{'-}(\theta) = g(\theta)g^\dagger (-\theta) P^{-}(\theta) \\[3mm]
& Q^{'+}(\theta) = f(\theta)g^\dagger (-\theta) Q^\dagger (\theta) \\[3mm]
& Q^{'-}(\theta) = g(\theta)f^\dagger (-\theta) Q^{-}(\theta).
\end{array}
\right.
$$

It is obvious that  $P^{\pm}(\theta)$ do not change under the 
transformations (28) and (29)
when $f(\theta)$ and $g(\theta)$ are both even functions of $\theta$. 
Moreover, $P^{\pm}(\theta)P^{\pm}(-\theta)$ and 
$Q^{\pm}(\theta)Q^{\mp}(-\theta)$ are invariant. 
If $g(\theta) = f(\theta) $, then $ R^{'}(\theta) = 
f(\theta)f^{\dagger}(-\theta) R(\theta) $, it is just the  \lq CDD ambiguity'.
Now we should see if the $ R(\theta) $ satisfies the Boundary Yang-Baxter 
equation , Unitarity condition and the Crossing Symmetry. It is easy to 
check that the $R(\theta)$ satisfies the Boundary Unitarity condition
\begin{equation}
R_a^{c}(\theta) R_{c}^{b}(-\theta) = \delta_a^{b},
\end{equation}
we found the $ R(\theta) $ satisfies the following Boundary Yang-Baxter 
equation
\begin{equation}
R_{a_2}^{c_2}(\lambda\theta_2)S_{a_1 c_2}^{c_1 
d_2}(\theta_1+\theta_2)R_{c_1}^{d_1}(\lambda\theta_1)S_{d_2 d_1}^{b_2 
b_1}(\theta_1 -\theta_2) = S_{a_1 a_2}^{d_1 
d_2}(\theta_1-\theta_2)R_{d_1}^{c_1}(\lambda\theta_1)S_{d_2 c_1}^{c_2 
b_1}(\theta_1 + \theta_2)R_{c_2}^{b_1}(\lambda\theta_2)
\end{equation}
where $S$ is the S matrix of Sine-Gordon model or Thirring model:
\begin{equation}
\left \{
\begin{array}{ll}
& S_{1 1}^{1 1}(\theta) = S_{2 2}^{2 2}(\theta) = \sin 
(\lambda(\pi + i\theta)) \rho(\theta)  \\[3mm]
& S_{1 2}^{1 2}(\theta) = S_{2 1}^{2 1}(\theta) = -\sin 
(i \lambda \theta ) \rho(\theta)  \\[3mm]
& S_{1 2}^{2 1}(\theta) = S_{2 1}^{1 2}(\theta) = \sin 
(\lambda(\pi ) \rho(\theta),  
\end{array}
\right.
\end{equation}
where $\rho(\theta)$ is
$$
\begin{array}{ll}
\rho(\theta) = - \frac{1}{\pi} \Gamma(\lambda) \Gamma(1 + 
\frac{i\lambda\theta}{\pi} ) \Gamma (1-\lambda - \frac{ i\lambda 
\theta}{\pi} ) 
\times \Pi^{\infty}_{l=1}\frac{F_l(-i\theta)F_l(\pi+i\theta)}{ F_l(0) 
F_l(\pi)} \\[3mm]
F_l(x) = \frac{\Gamma( 2l\lambda -\lambda x /\pi) \Gamma(1 +2l\lambda 
-\lambda x/ \pi )}{\Gamma( (2l + 1)\lambda -\lambda x /\pi) \Gamma(1 
+ (2l - 1) \lambda  - \lambda x /\pi)}.
\end{array}
$$

The Crossing Symmetry only make some constraints on the transformation 
factor $f(\theta)$ and $g(\theta)$. So we have obtained the  
reflection matrix $R_{\rm T}(\theta)$ for Thirring Model with the 
boundary (14),(15) and (19):
\begin{equation}
R_{\rm T}(\theta) = R(\lambda \theta) 
\end{equation}
which satisfy the boundary Y-B equation andthe C-U condition. Now we have 
to  compare this $R_{\rm T}$ matrix with $R_{\rm S}$
of \cite{GZ} using  the  invariant quantities\footnote{ It is difficult to 
compare directly the R matrices, because there is an ambiguity  on
both sides, and the results given in \cite{GZ} involve  infinite products of 
$\Gamma$-functions as follows:
$$ P^{\pm}(\theta) = \cos (\xi \pm \lambda u) R_s(u), ~~~~~~
Q^{\pm}(\theta) = \frac{k}{2}\sin(2\lambda u) R_s(u),
$$
where $u = -i\theta$,  $R_s(u) = R_0(u)R_1(u) $, $~~R_0(u) = 
\frac{F_0(u)}{F_0(-u)}$, $$
F_0(u) = \frac{\Gamma(1-2\lambda u/\pi)}{
\Gamma(\lambda-2\lambda u/\pi)} 
\times \Pi^{\infty}_{k=1}\frac{\Gamma(4\lambda k-2\lambda 
u/\pi)\Gamma(1+4\lambda k-2\lambda u/\pi) 
\Gamma(\lambda(4k+1))\Gamma(1+\lambda(4k-1))} 
{\Gamma(\lambda(4k+1)-2\lambda u/\pi) \Gamma(1+\lambda(4k-1)-2\lambda 
u/\pi) \Gamma(1+4\lambda k) \Gamma(4\lambda k)}, 
$$
$ R_1(u) = \frac{1}{\cos \xi}\sigma(\eta, u)\sigma( i \vartheta, u), $
where
$$
\sigma(x, u) =\frac{\Pi(x, \pi/2-u)\Pi(-x, \pi/2-u) 
\Pi(x, -\pi/2+u)\Pi(-x, -\pi/2+u)}{\Pi^2(x, \pi/2)\Pi^2(-x, \pi/2)}
$$
$$
\Pi(x, u) = \Pi^{\infty}_{l=0} \frac{\Gamma (1/2 + (2l + 1/2)\lambda + 
x/\pi -\lambda u/\pi ) \Gamma (1/2+(2l+3/2)\lambda + x/\pi)} { \Gamma(1/2 
+(2l+3/2)\lambda +x/\pi - \lambda u/\pi) \Gamma(1/2+(2l+1/2)\lambda +x/\pi)}
$$
$$
\sigma(x, u)\sigma(x, -u) = [\cos(x+\lambda u) \cos(x -\lambda 
u)]^{-1}\cos^2 x$$
there parameters $\eta$ and $\vartheta$ are determined through the 
equations
$$
\cos\eta \cosh { \vartheta} = - \frac{1}{k}\cos\xi, ~~~~\cos^2\eta + 
\cosh^2 \vartheta = 1+ \frac{1}{k^2}.
$$ }  to find if the Thirring model with this boundary is 
equivalent to the Sine-Gordon model. 

Indeed, the invariants  satisfy the following 
relations: 
\begin{equation}
P_{\rm T}^{\pm}(\theta)P_{\rm T}^{\pm}(-\theta) = \cos(\xi + i\lambda 
\theta)\cos(\xi - i\lambda \theta) R_s(\theta)R_s(-\theta)
\end{equation}
\begin{equation}
Q_{\rm T}^{\pm}(\theta)Q_{\rm T}^{\mp}(-\theta) = \frac{1}{4} 
k^2\sinh^2(2\lambda \theta)R_s(\theta)R_s(-\theta),
\end{equation}
where  $\xi$ and $k$ are two formal parameters involved 
in the results of \cite{GZ}, and   
\begin{equation}
R_s(\theta)R_s(-\theta) = [ \cos^2(\xi) + \sinh^2(\lambda \theta) + 
k^2\sinh^2(\lambda \theta)\cosh^2(\lambda \theta)]^{-1}.
\end{equation}

Inserting eq(27) into eq(36) and eq(37), one can get the relation between 
the boundary 
parameters $(h, \phi_0)$ and the formal parameters $(\xi, k)$ used in \cite{GZ}.
Thus,
\begin{equation}
\begin{array}{ll}
&[M\cosh(\lambda \theta) +  h\cosh(\lambda \theta +i\beta\phi_0)] 
[M\cosh(\lambda \theta) + 
{h}\cosh (\lambda \theta -i\beta\phi_0)]/(P_{\rm T}( 
\theta)P_{\rm T}(-\theta)) \\[3mm]
&= \cos(\xi -i\lambda\theta)\cos(\xi +i\lambda\theta)[\cos^2(\xi) + 
\sinh^2(\lambda \theta) + k^2\sinh^2(\lambda \theta)\cosh^2(\lambda 
\theta)]^{-1} \end{array} 
\end{equation}
and also,
\begin{equation}
\begin{array}{ll}
&M^2\sinh^2 (2\lambda \theta) /(P_{\rm T}( \theta)P_{\rm T}(-\theta)) 
\\[3mm] &= k^2\sinh^2(2\lambda \theta)[\cos^2(\xi) + 
\sinh^2(\lambda \theta) + k^2\sinh^2(\lambda \theta)\cosh^2(\lambda 
\theta)]^{-1}. \end{array} 
\end{equation}

For the fixed boundary case, $ h \rightarrow \infty $, we get $ k = 0 $ from 
eq(40), and eq(39) becomes an identity; it does not lead to any 
constraints on $\xi$. For the free boundary case, $h = 0 $,  we get 
$\cos^2 \xi = 1$ and $ k^2 = 1$ from eqs(39,40). It agrees with the 
results of\cite{GZ}  $ k= [\sin(\frac{\lambda\pi}{2})]^{-1} $ only when 
$\lambda = 2n+1, n= 0, \pm 1, \pm 2, ... $ .  
 This $ k $ was introduced 
in \cite{GZ} in order to get a pole of R matrix at $\theta = i\frac{\pi}{2}$
for free boundary. This pole can also be obtained  by putting a 
transformation factor such as $f(\theta) =g(\theta) $ and 
$f(\theta)f^\dagger (-\theta) = \frac{ s(h/M) +1-i\sinh \theta}{s(h/M) +1 
+ i\sinh \theta}$, where s is an arbitrary function satisfing  
$s(h/M)\Bigm|_{h = 0} = 0$ ( Such a \lq CDD ambiguity '
$\Phi(\theta)$ will not change any of the Boundary Yang-Baxter equations, 
the Unitary condition, or Crossing symmetry) which means we can not 
determine the $k$ only from the pole. However  the $ P^{\pm} (\theta)$ of 
\cite{GZ} tends to 
zero, but our $P_{\rm T}^{\pm}(\theta)$ does not when $\lambda = $ 
even integer .

For the general case,  we get
\begin{equation}
\left \{
\begin{array}{ll}
& k^2 = \frac{M^2}{M^2 + 2hM \cos (\beta\phi_0) + h^2} \\[3mm]
&\sin^2 (\xi) = \frac{h^2 \sin^2(\beta \phi_0)}{M^2 + 2hM \cos 
(\beta\phi_0)  + h^2}. 
\end{array}
\right.
\end{equation}

It  reproduces  the result of \cite{AKL} when $\lambda = 1 $.

\bigskip

{\bf Conclusion and Discussion} 

We have obtained the reflection matrix $R_{\rm T}(\theta)$ for the Thirring 
model with boundary (14), (15) and (19) 
which is regarded as the perturbation of the free boundary condition (15). 
This $R_{\rm T}(\theta)$  matrix is related directly to 
physical boundary parameters for the sine-Gordon model with integrable 
boundary (1) by using the relation between the sine-Gordon model and 
the Massive 
Thirring model.  This   $ R $-matrix  has  a degree of freedom  which is 
similar to the 
 \lq CDD ambiguity ', but its elements can be used to construct  invariant 
quantities so that we can compare  them with those of Ghoshal and 
Zamolodchikov as in eq(36) and eq(37). We found the simple boundary (14) 
for Thirring model is  equivalent to boundary (1) for Sine-Gordon model 
at least provided the coupling constant $\beta$ satisfies $ 
\frac{8\pi}{\beta^2} -1 = 2n+1 $.  

It should be noted that the soliton $\psi(x)$ and $A(\theta)$ in (21) are 
both fermionic operators. However, the Zamolodchikovs soliton operators
$A(\theta)$ are neither bosonic nor fermionic  in the general case. 
Only when $\lambda = 2n + 1$, can they  be regarded as fermionic 
operators. This is just why our $R_{\rm T}(\theta)$, which satisfies the  
boundary  Y-B equation and the C-U condition is equivalent to 
Ghoshal and Zamolodchikovs' only when $\lambda = 2n+1 $. We conjecture one 
can get 
$R_{\rm T}(\theta)$ which is equivalent to Ghoshal and Zamolodchikovs' for 
all $\lambda$ 
from equivalence between eq.(1) and eq.(14) when the $\psi(x)$ is 
expanded in  terms of Zamolodchikov's soliton operators. 

It is worth noting that while the boundary condition (2) is the unique 
form of integrable boundary condition for the sine-Gordon model, at least 
on the assumption its boundary potential ${\cal B}(\phi)$ only depends on 
$\phi$,  the boundary condition (20) is not the only integrable 
boundary condition for the massive Thirring model even if its boundary 
potential depends only on $\psi_i$ (and not its $x$-derivatives) at $ = 
0 $.
Moreover, there is other possibility leading to $j^0(x,t) \Bigm|_{x=0} =0 $
besides (12), i.e.  there are other free  boundary conditions for 
the Thirring model.  
The simplest case is ${\cal L}_{\rm T_{\rm free}} = {\cal L}_{\rm T_{\rm 
bulk}}$, but it is trivial\footnote{There is a similar case in Ising 
model\cite{GZ}. Recently, Mourad and Sasaki\cite{MS} also found  that the 
solutions become trivial when they use ${\cal L}_{\rm bulk}$ as free 
boundary Lagrangian for nonlinear sigma model on half plane. }.  Another 
non-trivial free boundary condition is 
$$
\begin{array}{l}
(1 + icM ) \psi_1(0) = a \psi_2(0) + b\psi^\dagger_2(0) - idM 
\psi^\dagger_1(0) \\[3mm]
 -icg\left (\psi^\dagger_1(0)\psi_1(0^-)+ \psi^\dagger_1(0^-)\psi_1(0) 
\right )\psi_2(0) 
 -idg\left (\psi^\dagger_1(0)\psi_1(0^-)+ \psi^\dagger_1(0^-)\psi_1(0) 
\right )\psi^\dagger_2(0).
\end{array}
$$
It is easy to check they lead to $j^0(x,t) \Bigm|_{x=0} =0 $ too, if 
$$
\left \{
\begin{array}{ll}
|c|^2 - |d|^2 = 0 \\ [3mm]
|b|^2 -|a|^2 = 1 +i M (c-c^{*}) \\[3mm]
ac^{*} -a^{*}c + bd^{*} - b^{*}d = 0.
\end{array}
\right.
$$ 
In other words,  the massive Thirring model with boundary is not 
equivalent  completely to the sine-Gordon model with boundary. The 
integrable  boundary condition for the 
massive Thirring model has more freedom than that for the sine-Gordon model. 
What is the general integrable condition of massive Thirring 
model? 
Obviously, there is much room for further development. 

\bigskip
\vfill
\eject
{\bf Acknowledgements}

Authors are  grateful to Ed. Corrigan for very useful discussions and 
suggestions as well as a critical reading of the original manuscript.
One of us (ZMS) would like also to thank R. Sasaki, J.M. Shen, R. Rietdijk,  
Y.X. Chen and J. 
Kim for useful discussions and the Mathematics Department of University of 
Durham for the kind hospitality. ZMS would like also to thank the  
Pao Foundation for a Fellowship and University of Durham for support.
HBG is supported by a K C Wong fellowship through Royal Society of, London.

\end{document}